\documentclass[pre,showpacs,preprint]{revtex4}

\usepackage{amsmath}

\usepackage{graphicx}
\usepackage{dcolumn}
\usepackage{bm}

\begin{document}

\title{ Power-law decay of the velocity autocorrelation function of a granular fluid in the homogeneous cooling state  }
\author{J. Javier Brey and M.J. Ruiz-Montero}
\affiliation{F\'{\i}sica Te\'{o}rica, Universidad de Sevilla,
Apartado de Correos 1065, E-41080, Sevilla, Spain}
\date{\today }

\begin{abstract}
The hydrodynamic part of the velocity autocorrelation function  of a granular fluid in the homogeneous cooling state has been calculated by using mode-coupling theory for a finite system with periodic boundary conditions. The existence of the shearing instability, leading to a divergent behavior of the velocity flow fluctuations, is taken into account.  A time region in which the velocity autocorrelation function  exhibits a power law decay, when time is measured by the number of collisions per particle,  has been been identified. Also the explicit form of the exponential asymptotic long time decay has been obtained.  The theoretical prediction for the power law decay is compared with molecular dynamics simulation results, and a good agreement is found, after taking into account finite size corrections.  The effects of approaching the shearing   instability are also explored.

\end{abstract}

\pacs{45.70.-n, 05.20.Dd,83.10.Pp}

\maketitle

\section{Introduction}
\label{s1}
Almost forty five years ago, Alder and Wainwright \cite{AyW70} reported, from a molecular dynamics simulation study,  the observation of an asymptotic power law decay of the velocity autocorrelation function (VACF) of a tagged particle in a fluid at equilibrium. At long times, the VACF decays as $t^{-d/2}$, where $d$ is the dimensionality of the system. They also proposed a simple hydrodynamic explanation, suggesting that the power law  decay is due to the slow relaxation of the velocity vortex that is generated by the motion of the tagged particle in the fluid. Theoretical analysis of this effect have been carried out since then using kinetic theory \cite{DyC70,Do75} and also more phenomenological mode coupling theories \cite{EHyvL70}. One of the main implications of the long time tails is that the time independent  Navier-Stokes transport coefficients, as defined by the Green-Kubo relations, do not exist in two-dimensional systems, outside the limit of dilute gases, where the long time-power law tails disappear. 

Granular gases have attracted a lot of attention in recent years, not only because of the rich phenomenoly they exhibit, but also because they are considered as a proving ground for kinetic theory and non-equilibrium statistical mechanics. The methods used for elastic molecular fluids have been extended to the case of particles colliding inelastically \cite{Du00,Go03}.   In particular, hydrodynamic equations to Navier-Stokes order have been derived with formal Green-Kubo like  expressions for the transport coefficients, both for dilute \cite{DyB02,BDyR03} and dense granular gases \cite{DByL02,BDyB08}. The low density expressions have been evaluated numerically by using the direct Monte Carlo simulation method \cite{ByR04a,ByR04b} and a satisfactory agreement has been found with the results obtained by  the Chapmann-Enskog procedure applied to the inelastic Boltzmann equation in the first Sonine approximation \cite{BDKyS98}. Note that in the low density limit considered in these works, algebraical decay of the correlation functions are not expected, by analogy with molecular gases. We are not aware of any numerical evaluation of the formal expressions of the transport coefficients for dense granular gases, aside from the self-diffusion coefficient in  a three dimensional system \cite{LByD02}. Consequently, it is not known whether the Green-Kubo like expressions for the transport coefficients of granular gases do actually exist beyond the low density limit in two-dimensional systems.

Due to dissipation in collisions, isolated  granular gases (or with periodic boundary conditions)  do not have an equilibrium Gibbs state, but rather there is a time-dependent homogeneous cooling state (HCS). The Navier-Stokes transport coefficients of a granular gas are expressed as averages over the distribution function corresponding to the HCS  \cite{DyB02,BDyR03,DByL02,BDyB08}. Consequently, it is a relevant question  to know whether time correlation functions computed in this state exhibit slow decaying long time tails.

Studying  the existence of hydrodynamics in a two-dimensional molecular system can appear as a rather formal and academic issue. Nevertheless, the situation is quite different when dealing with granular fluids. To reach and maintain them in a steady state, a permanent energy supply is necessary. The usual experimental procedures are either by means of an external field or by injecting energy through the boundaries. The price to be paid is that the system becomes highly inhomogeneous.  Recently, a new possibility is being explored. The idea is to place a granular gas between two large  parallel horizontal plates separated by a distance  larger than one particle diameter, but smaller than two, in such a way that the system is actually a granular monolayer \cite{OyU05,RPGRSCyM11}. To keep the system fluidized the horizontal plates are vibrated. Then, the two-dimensional dynamics obtained by projecting the motion of the grains on an horizontal plane is observed. Experiments show that the behavior of the projected system resembles that of a (two-dimensional) fluid. Developing a self-consistent hydrodynamics for it seems an interesting and promising topic. Of course, this requires the investigation of the decay of the fluctuations and correlations in the two-dimensional system. Actually, an experimental setup very similar to the one described above has been already used to measure velocity correlations on the hydrodynamic scale in a two-dimensional granular gas \cite{Puetal12}.

Long time tails in granular gases have already been studied. Kumaran \cite{Ku06} considered  sheared granular flows and reported the suppression of the power-law long time tail.  Also, the tail of the velocity correlation function has been analyzed in the stationary state generated by submitting the grains to an external noise force or thermostat \cite{FAyZ09}.  Although the results are interesting, it is worth to remark that the possible relationship of this kind of forces with real experiments has not been established, and that the results depend strongly on the specific form of the used  force \cite{FAyZ09}.
The case of an isolated granular gas has been investigated   by Ahmad and Puri \cite{AyP07}, by means of very extensive molecular dynamics simulations. Although their results for the velocity autocorrelation function in the two-dimensional case seem to suggest the existence of a power-law long time tail of exponent $-1$, when time is measured by the number of collisions per particle, they can not be considered as conclusive, and no comparison with theoretical predictions is carried out.  In ref.  \cite{HyO07},  several current autocorrelation functions have been investigated also in freely evolving granular fluids. Using a mode-coupling theory it is found, in particular, that the VACF has a long time decay of the same form as in molecular systems, when again  time is measured by the number of collisions per particle. Moreover, extensive molecular dynamics simulation results are presented, and it is claimed that they are consistent with the theory, although again no quantitative comparison is carried out, other than the identification of a  time  region in which the behavior of the correlation function follows the power law predicted by the theory for the asymptotic long time limit. In the last two studies,  very large systems are  considered, actually infinite in the theory developed in ref. \cite{HyO07},  so that the HCS is highly unstable. In practice, this means that the state of the system departs from the HCS very soon, developing velocity vortices and later on strong density inhomogeneities \cite{GyZ93,McyY94}. Consequently,  the physical meaning of the VACF becomes rather uncertain after a very short period of time. In particular, there is no reason to expect it to be related in a simple way with the self-diffusion coefficient of the simulated system, if this coefficient exists.

In this paper, the decay of the velocity autocorrelation function in a granular gas in the homogeneous cooling state (HCS) will be addressed. Both in the theory and in the simulations only systems in that state will be considered, although the effects of the shearing instability as it is approached will be taken into account. Therefore, in the systems analyzed here, the relationship between the VACF and the self-diffusion coefficient through a Green-Kubo formula is well established  \cite{DByL02}.  Although this implies the limitation to finite systems, it is possible to extrapolate the results and identify scaled behaviors that are independent from the size. In particular, this happens with the existence of a time window for which the VACF  has an algebraic decay that is correctly predicted by a mode-coupling theory, not only qualitatively but also quantitatively.

The remaining of this paper is organized as follows. In Sec. \ref{s2}, the definition of the self-diffusion coefficient in a granular gas in the HCS, and the steady-state representation of the latter, are summarized. It is important to stress that the existence of the steady representation is not an approximation, but an {\em exact} consequence of a change of variables. In particular, the self-diffusion coefficient is also determined by the VACF in the steady representation.  In Sec. \ref{s3},  the mode-coupling theory of Ernst, Hauge, and van  Leeuwen \cite{EHyvL70} is rederived  for a finite granular gas with periodic boundary conditions in the HCS. The peculiarity of the hydrodynamic fluctuations, playing a fundamental role in the theory, and the effects of the shearing instability, are analyzed in  Sec. \ref{s4}. The existence of a time scale over which the VACF has a power law decay in time is shown in Sec. \ref{s5}, where it is also discussed the exponential decay occurring in the asymptotic long time limit, due to the finite size of the system. In Sec. \ref{s6}, the method used to take into account finite size effects when comparing molecular dynamics simulation results and theoretical predictions is described. The power law tails from the simulations are identified in Sec. \ref{s7}. Both, the power law and its amplitude are shown to be in quite good agreement with the mode-coupling predictions.  Finally, Sec. \ref{s8} contains a short discussion of some relevant conceptual issues addressed in the paper, as well as some indications of possible extensions of the reported work.

\section{Self-diffusion coefficient in the steady-state representation of the homogeneous cooling state}
\label{s2}
In Ref. \cite{DByL02}, the self-diffusion coefficient $D(t)$ of a granular gas of $N$ inelastic hard spheres or disks of mass $m$ in the HCS is identified from the diffusion equation,
\begin{equation}
\label{2.1}
\frac{\partial}{\partial t} n_1({\bm r},t) -D(t) \nabla^2 n_1 ({\bm r},t) =0,
\end{equation}
where $n_1({\bm r},t)$ is the number  density of tagged particles at position ${\bm r}$ and time $t$. The formal expression of $D(t)$ is given in terms of the VACF by the Green-Kubo formula
\begin{equation}
\label{2.2}
D(t)= \frac{1}{d} \int_{0}^{t} dt^{\prime}\, <{\bm v}_{1}(t)  \cdot {\bm v}_{1}(t^{\prime});0>_{HCS}.
\end{equation}
Here, ${\bm v}_{1}(t)$ is the velocity of a  tagged particle at time $t$, and the angular brackets denote average with an initial distribution corresponding to the HCS of the system. Upon deriving the above expression, it is assumed that the system remains in the HCS in the time interval between $0$ and $t$.  Since all the particles of the system are mechanically equivalent, the VACF can be actually  computed by using any of the $N$ particles in the system. The HCS distribution function, $\rho_{HCS}$, giving the probability density for finding the particles at positions $\left\{ {\bm q}_{i} \right\}$  with velocities $\left\{ {\bm v}_{i} \right\}$, has the scaling form \cite{GyS95,BDyS97}
\begin{equation}
\label{2.3}
\rho_{HCS}\left( \left\{ {\bm q}_{i} \right\},\left\{ {\bm v}_{i} \right\},t  \right) = \left[ v_{0}(t)\right]^{-Nd} \rho^{*}_{HCS} \left( \left\{ {\bm q}_{ij} \right\},\left\{ \frac{{\bm v}_{i}}{v_{0}(t)} \right\} \right),
\end{equation}
where ${\bm q}_{ij} \equiv {\bm q}_{i}-{\bm q}_{j}$,
\begin{equation}
\label{2.4}
v_{0}(t) \equiv \left[ \frac{2T(t)}{m} \right]^{1/2},
\end{equation}
$T(t)$ is the granular temperature (the Boltzmann constant is set equal to unity upon defining the granular temperature from the average kinetic energy), and $\rho^{*}_{HCS}$ is a dimensionless isotropic distribution, which is invariant under space translations. In Eq. (\ref{2.3}), all the time dependence due to collisional cooling occurs through the granular temperature, that obeys the Haff law \cite{Ha83}
\begin{equation}
\label{2.5}
\frac{ \partial v_{0}(t)}{\partial t} =- \frac{1}{2}  \zeta (t) v_{0}(t).
\end{equation}
The cooling rate $\zeta (t)$ is proportional to $v_{0}(t)$. Then, it follows that the long time behavior of the thermal velocity $v_{0}(t)$  of a granular gas in the HCS is given by
\begin{equation}
\label{2.6}
v_{0}(t) \sim \left( \overline{\zeta} t \right)^{-1}.
\end{equation}
In this expression, a modified cooling rate coefficient,
\begin{equation}
\label{2.7}
\overline{\zeta} \equiv \frac{\zeta (t)}{ 2 v_{0}(t)},
\end{equation}
which does not depend on time, has been introduced.

It is convenient to change to  a new time scale in which the intrinsic time dependence of the HCS can be scaled out in some way. This allows to formally eliminate one of the two time explicit
dependencies of the VACF in Eq. (\ref{2.2}), making the theoretical analysis simpler and more direct \cite{DByL02}. Moreover, a direct molecular dynamics simulation of an freely evolving granular gas, as described by the dynamics in the actual phase space variables, has the limitation that the typical velocity of the particles
becomes very small rather soon and, therefore, numerical inaccuracies become very large. This is a very serious difficulty when the interest is focused on the long time behavior of a property of the system, as it is the case here.
To circumvent  this problem, the dynamics of a system of inelastic hard spheres or disks in the HCS is exactly mapped onto the dynamics around a steady state \cite{Lu01,BRyM04}. This is done by defining a new time scale $s$ by
\begin{equation}
\label{2.8}
\omega_{0}s = \ln \frac{t}{t_{0}},
\end{equation}
where $t_{0}$ and $\omega_{0}$ are two arbitrary constants. Consistently, the velocity ${\bm \omega}_{i}$ of a particle $i$ in the new time scale is given by
\begin{equation}
\label{2.9}
{\bm \omega}_{i}(t) \equiv \frac{ \partial {\bm q}_{i}}{\partial s} = \omega_{0} t {\bm v}_{i}(t).
\end{equation}
The particle dynamics consists now of an accelerated streaming between collisions,
\begin{equation}
\label{2.10}
\frac{\partial}{\partial s} {\bm q}_{i} = {\bm \omega}_{i} (s),
\end{equation}
\begin{equation}
\label{2.11}
\frac{\partial}{\partial s} {\bm \omega}_{i}(s) = \omega_{0} {\bm \omega}_{i}(s),
\end{equation}
while the collision rule in the new variables is the same as in the original ones, as  a consequence of the linearity of the transformation given by Eq. (\ref{2.9}) and the instantaneous character of collisions.

The distribution function of the HCS in the transformed phase space reads
\begin{equation}
\label{2.12}
\widetilde{\rho}_{HCS} \left( \left\{ {\bm q}_{i} \right\},\left\{ {\bm \omega}_{i} \right\},s \right) = \left[ v_{0}(t) \omega_{0} t\right]^{-Nd} \rho^{*}_{HCS} \left( \left\{ {\bm q}_{ij} \right\},\left\{ \frac{{\bm \omega}_{i}
}{v_{0}(t) \omega_{0} t} \right\} \right).
\end{equation}
In the long time limit, $v_{0}(t)$ is given by Eq. (\ref{2.6}) and, therefore, $\widetilde{\rho}_{HCS}$ becomes independent from the time $s$, and takes the stationary form
\begin{equation}
\label{2.13}
\widetilde{\rho}_{st} \left( \left\{ {\bm q}_{i} \right\},\left\{ {\bm \omega}_{i} \right\} \right)=  \left( \frac{\omega_{0}}{\overline{\zeta}} \right)^{-Nd} \rho^{*}_{HCS} \left( \left\{ {\bm q}_{ij} \right\},\left\{ \frac{{\bm \omega}_{i}
\overline{\zeta}}{ \omega_{0} } \right\} \right).
\end{equation}
Let us define the temperature $\widetilde{T}$ of the HCS  in the modified dynamics  by
\begin{equation}
\label{2.14}
\frac{d}{2} n \widetilde{T}(s) \equiv < \frac{m}{2}  \omega_{i}^{2};s >,
\end{equation}
where $n$ is the total number of particles density of the granular gas, and the angular brackets denote average in the HCS in the new phase space,
\begin{equation}
\label{2.15}
< A( \left\{ {\bm q}_{i} \right\}, \left\{ {\bm \omega}_{i} \right\} );s> \equiv \int \left( \prod_{i} d {\bm q}_{i} d {\bm \omega}_{i} \right) A(\left\{ {\bm q}_{i} \right\}, \left\{ {\bm \omega}_{i} \right\} ) \widetilde{\rho}_{HCS} \left( \left\{ {\bm q}_{i} \right\},\left\{ {\bm \omega}_{i} \right\},s \right).
\end{equation}
It follows from Eq.\  (\ref{2.13}) that, after some transient time interval, the system reaches a steady state with a temperature given by
\begin{equation}
\label{2.16}
\widetilde{T}_{st}= \frac{m}{2} \left( \frac{\omega_{0}}{ \overline{\zeta}} \right)^2.
\end{equation}
This relationship provides a very efficient way to measure the cooling rate of a granular gas in the  HCS by means of numerical  particle simulations using the steady representation \cite{DByL02,Lu01,BRyM04}.

The expression of the self-diffusion coefficient in the HCS, as given by Eq.\ (\ref{2.2}), becomes simpler in the steady representation, specially   when the initial time is chosen such that the asymptotic steady state has already been reached. Then, the Green-Kubo form for the self-diffusion coefficient in the steady representation  is trivially obtained,
\begin{equation}
\label{2.17} D(t) = \frac{1}{d} \left( \frac{T(t)}{\widetilde{T}_{st}} \right)^{1/2} \int_{0}^{s} d s ^{\prime}\, < {\bm \omega}_{1} (s^{\prime}) \cdot {\bm \omega}_{1}(0) >_{st}\ ,
\end{equation}
 where the ensemble average now is done with the stationary distribution reached with the modified dynamics in the long time limit. In the above equation, the two time scales $s$ and $t$ are related by Eq.\ (\ref{2.8}).

\section{Hydrodynamic component of the VACF}
\label{s3}
Equation (\ref{2.17}) shows that the relevant VACF for the calculation of the self-diffusion coefficient in the steady representation is
\begin{equation}
\label{3.1}
C_{\omega \omega}(s) \equiv \frac{1}{d} < {\bm \omega}_{1}(s) \cdot {\bm \omega}_{1}(0) >_{st}.
\end{equation}
In this function, the trivial dependence on time occurring in Eq. (\ref{2.2}) through the temperature of the HCS has been eliminated. In the following, attention will be focused in analyzing the long time behavior of $C_{\omega \omega}(s)$, that is important for  the calculation of the self-diffusion coefficient as well as  for the existence of the coefficient  itself.

The VACF  $C_{\omega \omega}(s)$ is a spatially homogeneous function, but it can be expressed as the volume integral of an inhomogeneous quantity,
\begin{equation}
\label{3.2}
C_{\omega \omega}(s)= \frac{1}{d} \int d{\bm r} \int d{\bm r}^{\prime}\,  < {\bm J}({\bm r},s) \cdot {\bm J}({\bm r}^{\prime},0) >_{st},
\end{equation}
where ${\bm J}$ is the microscopic current density of the tagged particle,
\begin{equation}
\label{3.3}
{\bm J}({\bm r},s) \equiv {\bm w}_{1} \delta \left[ {\bm r}-{\bm q}_{1}(s) \right].
\end{equation}
Next, following the ideas of Ernst, Hauge, and van Leeuwen \cite{EHyvL70,PyR75}, the ensemble average in Eq. (\ref{3.2}) will be performed in two steps. First, a partial average for fixed values of the ``relevant'' magnitudes is taken. Afterwards, the average over the fluctuations of these magnitudes is carried out. The relevant magnitudes in the present context are those which are coupled to the velocity of the tagged particle and relaxing slowly. Here, it will be assumed that they are the same for the dynamics of a tagged particle in a granular gas in the HCS as for self-diffusion in an elastic molecular fluid at equilibrium, namely the local number density of the tagged particle and the local momentum density. The reason is that they are conserved quantities and, therefore, are expected to decay on a slow  or hydrodynamic time scale.

The microscopic density of the tagged particle at point ${\bm r}$  is given by
\begin{equation}
\label{3.4}
N_{1} ({\bm r}) \equiv \delta( {\bm r}- {\bm q}_{1}).
\end{equation}
Instead of the local momentum density, the local velocity flow ${\bm W}({\bm r}) $ will be employed. It is defined as
\begin{equation}
\label{3.5}
{\bm W} ({\bm r}) \equiv \frac{1}{n} \sum_{j=1}^{N} {\bm \omega}_{j} \delta ({\bm r}- {\bm q}_{j}).
\end{equation}
In Fourier space, the above two quantities read
\begin{equation}
\label{3.6}
N_{1{\bm k}} = e^{i {\bm k} \cdot {\bm q}_{1}},
\end{equation}
\begin{equation}
\label{3.7}
{\bf W}_{\bm k}= \frac{1}{n} \sum_{j=1}^{N} {\bm \omega}_{j} e^{i {\bm k} \cdot {\bm q}_{j}}.
\end{equation}
A square ($d=2$) or cubic ($d=3$) system of side $L$ will be considered, and periodic boundary conditions employed.
Now, a constrained distribution in the modified phase space $\widetilde{\Gamma} \equiv \left\{ {\bm q}_{i},{\bm \omega}_{i}; i=1, \ldots,N \right\}$, for given macroscopic fields  $n_{1}$ and $\widetilde{\bm u}$, of the relevant magnitudes,  is defined as
\begin{equation}
\label{3.8}
\widetilde{\rho}^{(c)}_{st}  (\widetilde{\Gamma}; \left\{ n_{1{\bm k}}, \widetilde{\bm u}_{\bm k} \right\}) = \frac{\prod_{\bm k} \delta \left( n_{1 {\bm k}}- N_{1 {\bm k}} \right) \delta \left( \widetilde{\bm u}_{\bm k}- {\bm W}_{\bm k} \right) \widetilde{\rho}_{st}(\widetilde{ \Gamma})}{P( \left\{ n_{1 {\bm k}}, \widetilde{\bm u}_{\bm k} \right\})},
\end{equation}
where
\begin{equation}
\label{3.9}
P( \left\{ n_{1 {\bm k}}, \widetilde{\bm u}_{\bm k} \right\})= \int d \widetilde{\Gamma}\, \prod_{\bm k} \delta \left( n_{1 {\bm k}}- N_{1 {\bm k}} \right) \delta \left( \widetilde{\bm u}_{\bm k}- {\bm W}_{\bm k} \right) \widetilde{\rho}_{st}(\widetilde{ \Gamma}).
\end{equation}
The distribution $P( \left\{ n_{1 {\bm k}}, \widetilde{\bm u}_{\bm k} \right\})$ can be understood as the probability density of a fluctuation of both the number of tagged particles density and the local velocity fields. Note that the constrained distribution defined by Eq.\ (\ref{3.8}) is trivially normalized in the modified phase space.

The time dependence  in the VACF  $C_{\omega \omega}(s)$ can be made explicit by means of the pseudo-Liouville operator $ \widetilde{\cal{L}} $ of the granular system in the scaled dynamics \cite{DByL02}
\begin{equation}
\label{3.10}
C_{\omega \omega}(s)= \frac{1}{d} \int d {\bm r} \int d{\bm r}^{\prime} \int d \widetilde{\Gamma}\,   {\bm J}({\bm r}) \cdot  e^{-s \widetilde{\cal{L}}} \left[ {\bm J}({\bm r}^{\prime}) \widetilde{\rho}_{st} (\widetilde{\Gamma} ) \right].
\end{equation}
Using the constrained distribution defined by Eq.\ (\ref{3.8}) this can be rewritten as
\begin{eqnarray}
\label{3.11}
C_{\omega \omega}(s)& = &\frac{1}{d} \int \left( \prod_{\bm k} d n_{1 {\bm k}} d \widetilde{\bm u}_{\bm k} \right) P( \left\{ n_{1 {\bm k}}, \widetilde{\bm u}_{\bm k} \right\}) \nonumber \\
&& \times \int d {\bm r} \int d{\bm r}^{\prime} \int d \widetilde{\Gamma}\,  {\bm J}({\bm r}) \cdot  e^{-s \widetilde{\cal{L}}} \left[ {\bm J}({\bm r}^{\prime}) \widetilde{\rho}^{(c)}_{st}  (\widetilde{\Gamma}; \left\{ n_{1{\bm k}}, \widetilde{\bm u}_{\bm k} \right\}) \right].
\end{eqnarray}
Consider the average of the current density in the restricted steady HCS ensemble,
\begin{equation}
\label{3.12}
{\bm j}^{(c)} ({\bm r}) \equiv \int d \widetilde{\Gamma} {\bm J} ({\bm r}) \widetilde{\rho}^{(c)}_{st}  (\widetilde{\Gamma}; \left\{ n_{1{\bm k}}, \widetilde{\bm u}_{\bm k} \right\}).
\end{equation}
It seems sensible to assume that
\begin{equation}
\label{3.13}
{\bm j}^{(c)} ({\bm r}) =n_{1}({\bm r}) \widetilde{\bm u}({\bm r}),
\end{equation}
where $n_{1}({\bm r}) $ and $ \widetilde{\bm u}({\bm r})$ are the inverse Fourier transformed of $n_{1 {\bm k}}$ and $\widetilde{\bm u}_{\bm k}$, respectively. Then, using the Parceval relation, it is found
\begin{equation}
\label{3.14}
\int d{\bm r} {\bm j}^{(c)} ({\bm r})=\frac{1}{V} \sum_{\bm k} n_{1 {\bm k}} \widetilde{\bm u}_{\bm -k},
\end{equation}
with $V=L^{d}$ the volume of the system. The crucial hypothesis of the theory will be introduced at this point. The right hand side of Eq.\  (\ref{3.11}) is split into a fast, kinetic relaxation towards a local steady distribution followed by a much slower relaxation controlled by hydrodynamics,
\begin{equation}
\label{3.15}
C_{\omega \omega}(s) = C_{\omega \omega, \text{kin}} (s) + C_{\omega \omega, \text{hyd}}(s).
\end{equation}
The concept of local steady state is a direct extension of the widely used local equilibrium state of molecular systems. It is a reference state in which the system is considered to be locally in the steady HCS, but with the hydrodynamic fields density of tagged particle and flow velocity, being functions of space and time \cite{DByB08}. In the following, attention will be restricted to the second term on the right hand side of the above equation, that is expected to dominate for $s \gg s_{\text rel}$, where $s_{\text rel}$ is some characteristic microscopic relaxation time, i.e.
\begin{equation}
\label{3.16}
C_{\omega \omega}(s) \simeq C_{\omega \omega, \text{hyd}}(s),
\end{equation}
for $s \gg  s_{\text rel}$. Moreover, in the spirit of the above time scale separation, it is assumed that in the hydrodynamic regime,
\begin{equation}
\label{3.17}
e^{-s \widetilde{\cal{L}}} \left[ {\bm J}({\bm r}^{\prime}) \widetilde{\rho}^{(c)}_{st}  (\widetilde{\Gamma}; \left\{ n_{1{\bm k}}, \widetilde{\bm u}_{\bm k} \right\}) \right] \simeq {\bm j}^{(c)} ({\bm r}^{\prime}) e^{-s \widetilde{\cal{L}}}  \widetilde{\rho}^{(c)}_{st}  (\widetilde{\Gamma}; \left\{ n_{1{\bm k}}, \widetilde{\bm u}_{\bm k} \right\}).
\end{equation}
Substitution of this expression into Eq.\ (\ref{3.11}), and use of Eq.\ (\ref{3.14}), yields
\begin{eqnarray}
\label{3.18}
C_{\omega \omega, \text{hyd}}(s) &  \simeq & \frac{1}{Vd} \sum_{{\bm k}_{1}} \int \left( \prod_{\bm k} dn_{1 {\bm k}} d \widetilde{\bm u}_{\bm k} \right) P( \left\{ n_{1 {\bm k}}, \widetilde{\bm u}_{\bm k} \right\})\nonumber \\
& & \times  n_{1 {\bm k}_{1}} \widetilde{\bm u}_{- {\bm k}_{1}} \cdot \int d{\bm r} \int d \widetilde{\Gamma}  {\bm J}({\bm r}) e^{-s \widetilde{\cal{L}}}  \widetilde{\rho}^{(c)}_{st}  (\widetilde{\Gamma}; \left\{ n_{1{\bm k}}, \widetilde{\bm u}_{\bm k} \right\}).
\end{eqnarray}
The next approximation is based once again on the same physical picture. Since the density of the tagged particle and the local velocity flow evolve very slowly for small values of the wavenumber vector, the distribution function $\widetilde{\rho}_{st}^{(c)}$ adjusts itself continuously, remaining with the same functional form on the long time scale,
\begin{equation}
\label{3.19}
e^{-s \widetilde{\cal{L}}}  \widetilde{\rho}^{(c)}_{st}  (\widetilde{\Gamma}; \left\{ n_{1{\bm k}}, \widetilde{\bm u}_{\bm k} \right\}) \simeq \widetilde{\rho}^{(c)}_{st}  (\widetilde{\Gamma}; \left\{ n_{1{\bm k}} (s), \widetilde{\bm u}_{\bm k}(s) \right\}).
 \end{equation}
Moreover, $ n_{1{\bm k}} (s)$ and $ \widetilde{\bm u}_{\bm k}(s)$ are determined  by the linearized hydrodynamic equations (to Navier-Stokes order) with the initial conditions  $n_{1{\bm k}} $ and $ \widetilde{\bm u}_{\bm k}$. This seems legitimate for small enough wave-vectors ${\bm k}$. Thus it must be verified a posteriori whether the long time behavior of $C_{\omega \omega}$ is actually governed by small wave numbers.
When Eq. (\ref{3.19}) is substituted into Eq.\ (\ref{3.18}) one gets
\begin{equation}
\label{3.20}
C_{\omega \omega, \text{hyd}}(s)   \simeq  \frac{1}{Vd} \sum_{{\bm k}_{1}} \int \left( \prod_{\bm k} dn_{1 {\bm k}} d \widetilde{\bm u}_{\bm k} \right) P( \left\{ n_{1 {\bm k}}, \widetilde{\bm u}_{\bm k} \right\}) n_{1 {\bm k}_{1}} \widetilde{\bm u}_{- {\bm k}_{1}}  \cdot \int d{\bm r}\,  {\bm j}^{(c)} ({\bm r},s),
\end{equation}
where Eq.\ (\ref{3.12}) has been employed. If now Eq.\ (\ref{3.14}) is taken into account, the above expression can be rewritten as
\begin{equation}
\label{3.21}
C_{\omega \omega, \text{hyd}}(s)   \simeq  \frac{1}{V^{2} d} \sum_{{\bm k}_{1}}         \sum_{{\bm k}_{2}}         \int \left( \prod_{\bm k} dn_{1 {\bm k}} d \widetilde{\bm u}_{\bm k} \right) P( \left\{ n_{1 {\bm k}}, \widetilde{\bm u}_{\bm k} \right\}) n_{1 {\bm k}_{1}}  \cdot \widetilde{\bm u}_{- {\bm k}_{1}}  n_{1 {\bm k}_{2}}(s) \widetilde{\bm u}_{- {\bm k}_{2}} (s).
\end{equation}
In linear hydrodynamics around the HCS, the fluctuations of the density of the tagged particle and of the velocity flow are not coupled, i.e. they are statistically independent. Moreover, different wave-vectors ${\bm k}$ are also uncoupled, so that
\begin{equation}
\label{3.22}
P( \left\{ n_{1 {\bm k}}, \widetilde{\bm u}_{\bm k} \right\}) = \prod_{\bm k} P_{n_{1}}(n_{1 \bm k}) P_{\widetilde{\bm u}_{\bm k}}(\widetilde{{\bm u}}_{\bm k}),
\end{equation}
where the marginal probability distributions $P_{n_{1}}$ and $P_{\bm u_{\bm k}}$ verify the normalization conditions
\begin{equation}
\label{3.23}
\int dn_{1 {\bm k}}\,  P_{n_{1}}(n_{1 \bm k}) = \int d \widetilde{\bm u}_{\bm k}\, P_{\widetilde{\bm u}_{\bm k}}(\widetilde{\bm u}_{\bm k}) =1.
\end{equation}
In addition, the isotropy of fluctuations in the HCS implies that the only non-vanishing contributions in Eq.\ (\ref{3.21}) are those with ${\bm k}_{1}=- {\bm k}_{2}$. In this way, it is obtained
\begin{equation}
\label{3.24}
C_{\omega \omega, \text{hyd}}(s)   \simeq  \frac{1}{V^{2} d} \sum_{\bm k}         \int  dn_{1 {\bm k}}  \int d \widetilde{\bm u}_{\bm k} P_{n_{1}}(n_{1 \bm k}) P_{\widetilde{\bm u}_{\bm k}}(\widetilde{\bm u}_{\bm k}) n_{1 {\bm k}} n_{1- {\bm k}} (s) \widetilde{\bm u}_{\bm k} \cdot \widetilde{\bm u}_{-{\bm k}}(s).
\end{equation}

\section{Hydrodynamic fluctuations}
\label{s4}

 As indicated above, the linearized hydrodynamic equations for a granular fluid will be  used to evaluate the time evolution  of the fluctuations of the hydrodynamic fields. The diffusion equation (\ref{2.1}) in the $s$-time scale has the form
 \begin{equation}
 \label{4.1}
 \left( \frac{\partial}{\partial s}+ \widetilde{D} k^{2} \right) n_{1 {\bm k}}(s) =0,
 \end{equation}
 with
 \begin{equation}
 \label{4.2}
 \widetilde{D}= \int_{0}^{\infty} ds\, C_{\omega \omega} (s).
 \end{equation}
Here it has been assumed that the VACF decays fast enough so the time integral on the right hand side of the above equation exists, leading to a time-independent transport coefficient $\widetilde{D}$. Integration of Eq.\ (\ref{4.1}) gives
\begin{equation}
\label{4.3}
n_{1-{\bm k}} (s) = n_{1- {\bm k}}(0) e^{-k^{2} \widetilde{D} s}.
\end{equation}
 The flow velocity fluctuations $\widetilde{\bm u}_{\bm k}$ can be decomposed into their longitudinal and transversal components, $\widetilde{\bm u}_{{\bm k} \parallel}$ and   $\widetilde{\bm u}_{{\bm k} \perp}$, defined by
\begin{equation}
 \label{4.4}
 \widetilde{\bm u}_{{\bm k} \parallel} = \frac{\widetilde{\bm u}_{\bm k} \cdot {\bm k}}{k^{2}} \, {\bm k},
 \end{equation}
\begin{equation}
 \label{4.5}
 \widetilde{\bm u}_{{\bm k} \perp} = \widetilde{\bm u}_{\bm k}- \frac{\widetilde{\bm u}_{\bm k} \cdot {\bm k}}{k^{2}} \, {\bm k},
 \end{equation}
 respectively. In the new time scale $s$, $\widetilde{\bm u}_{\perp}$ obeys the closed equation \cite{BDKyS98}
 \begin{equation}
 \label{4.6}
 \left( \frac{\partial }{\partial s}- \omega_{0}+ \widetilde{\eta} k^{2} \right) \widetilde{\bm u}_{{\bm k} \perp}(s)=0,
 \end{equation}
 where $\widetilde{\eta}$ is defined from the shear viscosity $\eta$ by
 \begin{equation}
 \label{4.7}
 \widetilde{\eta}= \frac{\eta}{m n} \left( \frac{\widetilde{T}_{st}}{T(t)} \right)^{1/2}\, .
 \end{equation}
 The solution of Eq. (\ref{4.6}) is
 \begin{equation}
 \label{4.8}
  \widetilde{\bm u}_{{\bm k} \perp}(s)=  \widetilde{\bm u}_{{\bm k} \perp} e^{ ( \omega_{0}-\widetilde{\eta} k^{2})s}.
  \end{equation}
 In this expression, it is manifest that  perturbations of the modified transversal velocity grow in time for small wave vectors, i.e. in large
 enough systems. This is the origin of the shearing instability of the HCS \cite{GyZ93,McyY94}, in which strong density inhomogeneities are developed. They are generated  by  nonlinear coupling contributions of the transversal velocity \cite{BRyC99}. To avoid misunderstandings, it is important to realize that the linear stability criterion for $\widetilde{\bm u}_{{\bm k} \perp}$ following from Eq. (\ref{4.8}) does not depend on the arbitrary value of $\omega_{0}$.  The critical value of the wavenumber vector $k_{\perp}$, such that  transversal modes with $ k <k_{\perp}$ are unstable, is given by the solution of the equation
 \begin{equation}
 \label{4.9}
 \omega_{0}- \widetilde{\eta} k_{\perp}^{2}=0,
 \end{equation}
that, using Eq. (\ref{2.16}), is seen to be equivalent to
\begin{equation}
\label{4.10}
\overline{\zeta} -\left( \frac{1}{2 m T(t)} \right)^{1/2} \eta k_{\perp}^{2}=0.
\end{equation}
It is now clear that $k_{\perp}$ does not depend on  $\omega_{0}$ or $T(t)$, since $\eta$ is proportional to $T(t)^{1/2}$.

Now, the contributions to $C_{\omega \omega, \text{hyd}}(s)$ from the fluctuations of the longitudinal component of the flow field $\widetilde{\bm u}_{{\bm k} \parallel} $ should be considered. Determining its time evolution using the linear hydrodynamic equations for a granular fluid is a rather involved problem, since  $\widetilde{\bm u}_{{\bm k} \parallel} $ can not be associated to a unique hydrodynamic mode, and its functional form changes depending on the range of values of the wavenumber considered \cite{BDKyS98}. Here the focus will be on small systems, in which the allowed values of $k$, i.e. compatible with the periodic boundary conditions, are such that the system exhibits a sound-like hydrodynamic mode, at least in dilute systems \cite{BDKyS98}. Then, it will be assumed that the contribution from the longitudinal component of the flow velocity to $C_{\omega \omega}(s)$ for large $s$ is sub-dominant, as it is the case in systems of elastic particles \cite{EHyvL70,PyR75}. Alternatively, neglecting the contributions from the longitudinal component of $\widetilde{\bm u}$, can be considered as an incompressible fluid approximation.

Substitution of Eqs. (\ref{4.3}) and (\ref{4.8}) into Eq.\ (\ref{3.24}) and use of
\begin{equation}
\label{4.11}
\int dn_{1 {\bm k}}\, P(n_{1 {\bm k}}) |n_{1 {\bm k}}|^{2}=1,
\end{equation}
leads to
\begin{equation}
\label{4.12}
C_{\omega \omega, \text{hyd}} (s) \simeq \frac{ e^{\omega_{0} s}}{V d} \sum_{\bm k} A({\bm k}) e^{-k^2s \left( \widetilde{\eta}+ \widetilde{D} \right)}.
\end{equation}
where
\begin{equation}
\label{4.13}
A({\bm k}) = \frac{1}{V} \int d \widetilde{\bm u}_{\bm k} P (\widetilde{\bm u}_{\bm k})\,  |\widetilde{\bm u}_{\bm k}|^{2}.
\end{equation}
In Eq.\ (\ref{4.12}) the range of values of ${\bm k}$ is restricted by $ k_{m} \leq k \leq k_{M}$, where $k_{m}$ is set by the system size, $k_{m} = 2 \pi /L$, and $k_{M}$ is of the order of $2\pi$ times the inverse of the mean free path. This upper bound condition is needed to guarantee that the local HCS varies slowly in space.

In elastic, molecular fluids in thermal equilibrium, the calculation of the second moment of the velocity field fluctuations is trivial. However, in a granular gas in the HCS,  velocity correlations are present, rendering the computation of $A({\bm k})$ more difficult. Using inelastic fluctuating hydrodynamics, van Noije {\em et al.} \cite{vNEByO97} computed the quantity
\begin{equation}
\label{4.14}
S_{\perp}(k,t) \equiv \frac{1}{V (d-1)} < {\bm u}_{{\bm k} \perp}(t) \cdot {\bm u}_{{\bm k} \perp} (t);0>_{HCS}.
\end{equation}
Here ${\bm u}$ is the local flow velocity fluctuation in the original time scale $t$. The result, in the notation used in this paper is
\begin{equation}
\label{4.15}
S_{\perp}(k,t)= \frac{T(t)}{m n}\left[ 1+ \frac{e^{2 \omega_{0} s(1-k^{2} /k_{\perp}^{2})}-1}{1-k^{2} /k_{\perp}^{2}} \right].
\end{equation}
An equivalent result was obtained by using a single relaxation model  kinetic theory \cite{BMyR98}. Moreover, in recent years, a description of fluctuating hydrodynamics in dilute granular gases more rigorous and complete than the one used in ref. \cite{vNEByO97} has been developed \cite{BMyG09}. Nevertheless,  for the values of the coefficient of normal restitution to be considered here, that correspond to the quasi-elastic limit,  both theories lead to results quantitatively indistinguishable for $S_{\perp}(k,t)$. For systems in which the HCS is stable, it is $k_{m} > k_{\perp}$, and the long time limit of the above expression is
\begin{equation}
\label{4.16}
S_{\perp}(k,t)= \frac{T(t)}{mn}\, \frac{k^{2}}{k^{2}-k_{\perp}^{2}}.
\end{equation}
It is worth to point out that the result given by this equation has the scaling property implied by the assumed form of the distribution function of the HCS, Eq.\ (\ref{2.3}), while Eq.\ (\ref{4.15}) does not. Then, taking into account the relationship between the velocity fields in the time scales $t$ and $s$, it is identified
\begin{equation}
\label{4.17}
A({\bm k})= \frac{(d-1) \widetilde{T}_{st}}{mn}\, \frac{k^{2}}{k^{2}-k_{\perp}^{2}}\, .
\end{equation}
Introduction of this into Eq.\ (\ref{4.12}) provides the explicit expression for the  hydrodynamic part of the steady VACF of a granular gas in the HCS,
\begin{equation}
\label{4.18}
C_{\omega \omega, \text{hyd}} (s) \simeq \frac{(d-1) \widetilde{T}_{st}e^{\omega_{0} s}}{mnVd}  \sum_{\bm k}\nolimits  ^{(\prime)} \frac{k^{2} e^{-k^2s \left( \widetilde{\eta}+ \widetilde{D} \right)}}{k^{2}-k_{\perp}^{2}}\, .
\end{equation}
The prime in the sum over ${\bm k}$ indicates the restriction imposed by the two cutoffs mentioned before. This expression leads to an exponentially increasing VACF on the time scale $s$ for values of $k < k_{c} \equiv \omega_{0}/(\widetilde{\eta}+ \widetilde{D})$. Nevertheless, this is not physically relevant since the HCS is unstable in systems large enough as to allow wave-vectors with these values. This is because $k_{c} < k_{\perp} $, where $k_{\perp}$, defined by Eq. (\ref{4.9}), is the lower bound of the wave-vector for the stability of the HCS with regards to the shearing instability. Moreover,  in Eq.\, (\ref{4.18}) $C_{\omega \omega, \text{hyd}} (s)$ also exhibits a divergent amplitude when $k_{m}$ approaches $k_{\perp}$. This divergence is just the shearing instability, and clearly implies a divergent behavior of the self-diffusion coefficient as given by the Green-Kubo expression, Eq. (\ref{2.2}), as the instability is approached.

\section{Two limiting time regimes of the VACF}
\label{s5}
In this section, two particular limits of Eq.\ (\ref{4.18}) will be investigated. Suppose first that $L \ll L_{c}$, where $L_{c}$ is the critical size of the system for the shearing instability, i.e. (see Eq.\ (\ref{4.10}))
\begin{equation}
\label{5.1}
L_{c} = 2 \pi \left( \frac{\widetilde{\eta}}{\widetilde{v}_{st} \overline{\zeta}} \right)^{1/2}\, ,
\end{equation}
where $\widetilde{v}_{st} \equiv \left( 2 \widetilde{T}_{st}
 /m \right)^{1/2}$. Then it is,
\begin{equation}
\label{5.2}
\left( \frac{k}{k_{\perp}} \right)^{2} \geq \left( \frac{2 \pi}{L k_{\perp}} \right)^2  \gg1,
\end{equation}
for all the allowed values of $k$. It follows that  Eq.\ (\ref{4.18}) can be approximated by
\begin{equation}
\label{5.3}
C_{\omega \omega, \text{hyd}} (s) \simeq \frac{(d-1) \widetilde{T}_{st}e^{\omega_{0} s}}{mnVd}  \sum \nolimits _{\bm k}  ^{(\prime)}  e^{-k^2s \left( \widetilde{\eta}+ \widetilde{D}
\right)}.
\end{equation}
A useful representation of this expression can be obtained by means of  the {\em d}-dimensional Poisson-sum formula \cite{Br58}
\begin{equation}
\label{5.4}
L^{-d} \sum_{\bm n} g \left( \frac{n}{L} \right) = \sum_{\bm l} \int d{\bm r}\, e^{-2 \pi i L {\bm l} \cdot {\bm r}} g(r),
\end{equation}
where ${\bm n}$ and ${\bm l}$ are {\em d}-dimensional vectors whose components are integers, the summations extend from $- \infty$ to $+ \infty$ for each of the components of ${\bm n}$ and ${\bm l}$, and the ${\bm r}$ integration extends over the infinite {\em d}-dimensional space. Indeed, use of Eq. (\ref{5.4}) into Eq.\ (\ref{5.3}) yields
\begin{equation}
\label{5.5}
C_{\omega \omega, \text{hyd}} (s) \simeq  \frac{(d-1) \widetilde{T}_{st}e^{\omega_{0} s}}{mnd} \left[ \frac{1}{4 \pi \left( \widetilde{\eta} + \widetilde{D} \right)s} \right]^{d/2} \sum_{\bm l} e^{- \frac{l^{2} L^{2}}{4 (\widetilde{\eta}+ \widetilde{D} )s}}.
\end{equation}
Consider times $s$ such that
\begin{equation}
\label{5.6}
s \ll s_{0} = \frac{L^{2}}{4 \left( \widetilde{\eta} + \widetilde{D} \right) \pi^{2}}\, .
\end{equation}
Taking into account that its has been supposed that $L \ll L_{c}$ and Eq.\ (\ref{4.9}), it is easy to verify that the above condition also implies that
\begin{equation}
\label{5.7}
\omega_{0} s \ll 1.
\end{equation}
As a consequence, Eq. \ (\ref{5.5}) reduces to
\begin{equation}
\label{5.8}
C_{\omega \omega, \text{hyd}} (s) \simeq  \frac{(d-1) \widetilde{T}_{st}}{mnd} \left[ \frac{1}{4 \pi \left( \widetilde{\eta} + \widetilde{D} \right)s} \right]^{d/2}.
\end{equation}
This prediction is expected to be valid for short times on the hydrodynamic scale, in the sense of being $s \ll s_{0}$, but, on the other hand, the equation only holds after all the fast non-hydrodynamic modes have decayed. It is important to compare this result with the  long time tails the of the VACF  in an equilibrium  molecular system. Formally, the derived expressions in both cases look the same \cite{EHyvL70,PyR75}, but there are relevant conceptual differences. The time scale $s$ used in Eq. (\ref{5.8}) is related with the original time scale  $t$ by  Eq.\ (\ref{2.8}), so that the algebraic $s^{-1}$ decay transforms into an even slower logarithmic decay on the  time scale $t$. Actually, the time scale $s$ is proportional to the cumulated number of collisions per particle for a system of inelastic hard spheres or disks in the HCS \cite{DByL02}. In a molecular system in equilibrium, this number is just proportional to $t$. Another, more significant, difference is that Eq.\ (\ref{5.8}) has been obtained here as valid in an intermediate time regime, while the corresponding expression in a molecular system has been  usually derived as valid for asymptotically long times or, more precisely, for all the hydrodynamic decay of the VACF. The reason for this strong difference is that in molecular systems, the thermodynamic limit was considered, while in the present case such a limit can not be taken, at least in the usual way,  for the HCS due to the shearing instability. A particularly enlightening discussion of the finite size effects in the evaluation of the VACF in a molecular system can be found in ref. \cite{EyW82}. A final comment on Eq.\ (\ref{5.8}) seems appropriate. Although the time scale  in which the behavior  predicted by this equation holds depends on the size $L$ of the system, the shape of the VACF in that time region is independent of $L$, as long as it is well inside the range in which the HCS is stable with regards to the shearing instability.

The second relevant limit of Eq.\ (\ref{4.18}) to be considered, is its asymptotic behavior for very large times. Then, it will be supposed now that $s \gg s_{0}$, with $s_{0}$ given by Eq.\ (\ref{5.6}). In this limit, it is evident that contributions with the smallest possible value of $k$, $k_{m}$, will dominate in Eq. (\ref{4.18}). Taking into account that, in a square or cubic geometry with periodic boundary conditions  there are $d$ modes with $k=k_{m}$, one gets
\begin{equation}
\label{5.9}
C_{\omega \omega, \text{hyd}} (s) \simeq \frac{(d-1) \widetilde{T}_{st}}{mnV}   \frac{k_{m}^{2} e^{-\left[ k_{m}^2 \left(\widetilde{\eta}+ \widetilde{D} \right)-\omega_{0} \right]s}}{k_{m}^{2}-k_{\perp}^{2}}\, .
\end{equation}
Note that the stability condition for the HCS, $\widetilde{\eta} k_{m}^2 > \omega_{0}$, implies that $C_{\omega \omega, \text{hyd}} (s)$ always decays exponentially in the asymptotic long time limit in the scale $s$. On the other hand, the amplitude of the decay diverges as the instability is approached.

\section{Molecular dynamics simulation results for the VACF}
\label{s6}
In order to check the accuracy of the theory developed in the previous sections, event driven molecular dynamics (MD) simulations of a system of inelastic hard disks have been performed.  The coefficient of restitution has been always taken as $\alpha=0.99$, while the number density has been varied in the interval  $ 0.231 \leq n \sigma^{2} \leq 0.385$. The maximum and minimum density values in the interval correspond to densities studied by Alder and Wainwright in their seminal paper \cite{AyW70}, so it is possible a direct comparison with their results for elastic systems. Simulations with different numbers of particles in the range $500 \leq N \leq 1000$ have been carried out for each density, modifying accordingly the size $L$ of the system. By estimating the  critical wave number for the shearing instability of the HCS, $k_{\perp}$,  using the transport coefficients and the cooling rate obtained by means of  the revised Enskog theory for inelastic hard disks \cite{GSyM07}, it is found that in all the simulations considered, the minimum wave number allowed, $k_{m}$, is significantly larger than $k_{\perp}$. Moreover, it was always checked in the simulations that the system remained homogeneous and with no appreciable velocity vortices. This latter condition is specially relevant in the present context, since the fluctuations of  the transversal velocity play a dominant role in the long time, hydrodynamic behavior of the VACF. Attention has also been paid to consider times such that the particles do not interact with their images in other cells. To avoid it, Erpenbeck and Wood \cite{EyW82} suggested to consider times smaller than the time a sound wave takes to travel a distance equal to the linear size of the system $L$.  This criterium is adopted in most studies of the VACF, see for instance \cite{DOyL06}, and has also been used here.

In the simulations, the steady representation of the HCS discussed in Sec.\ \ref{s2} has been used. Although it is not relevant for the results presented below, let us mention that the value of the parameter $\omega_{0}$ was chosen such that, if the Enskog theory were exact, the value of the steady temperature, as predicted by Eq.\ (\ref{2.16}), would be the same as the initial one $\widetilde{T}_{st}=\widetilde{T}(0)$. In all the simulations, the applied procedure was as follows. The system was prepared in an initial spatially homogeneous state, with a Gaussian velocity distribution. This initial state was allowed to evolve with the modified dynamics and periodic boundary conditions,  until a steady state was reached. Then, the VACF was measured. The results were averaged over several trajectories, typically a few hundreds.

As already pointed out, the prediction about  the existence of a hydrodynamic period over which the VACF exhibits a power law decay in the $s$ time scale, and the specific form of this decay, do not involve the size of the system, aside from requiring it to be smaller  than the critical size for the shearing instability, as it can be seen in Eq.\ (\ref{5.8}). Nevertheless, when computing the VACF from MD simulations, there are two main (related) reasons  for which  the finite size of the system must be taken into account to compare numerical results  with the theoretical predictions. The first reason is due to conservation of the total momentum of the system \cite{TKTAEyF90}, that implies that a given particle is actually moving in a fluid with a non-zero average velocity. The second cause is the difference between the ensemble in which the theory is developed and the one used in the simulations \cite{EyW82}. Both effects have been analyzed and quantified for molecular systems in equilibrium, but the arguments used in this case are not easy to extend to the present intrinsic non-equilibrium system with the particles submitted to an effective acceleration. Then, an heuristic view will be adopted. It will be assumed, as it is the case in elastic systems, that both effects scale with $N^{-1}$ to leading order, and the proportionality constant will be determined from the simulations themselves. Of course, the simulations also provide a test, {\em a posteriori}, of whether the leading dependence on $N$ is indeed the assumed one.

\begin{figure}
\includegraphics[scale=0.6,angle=0]{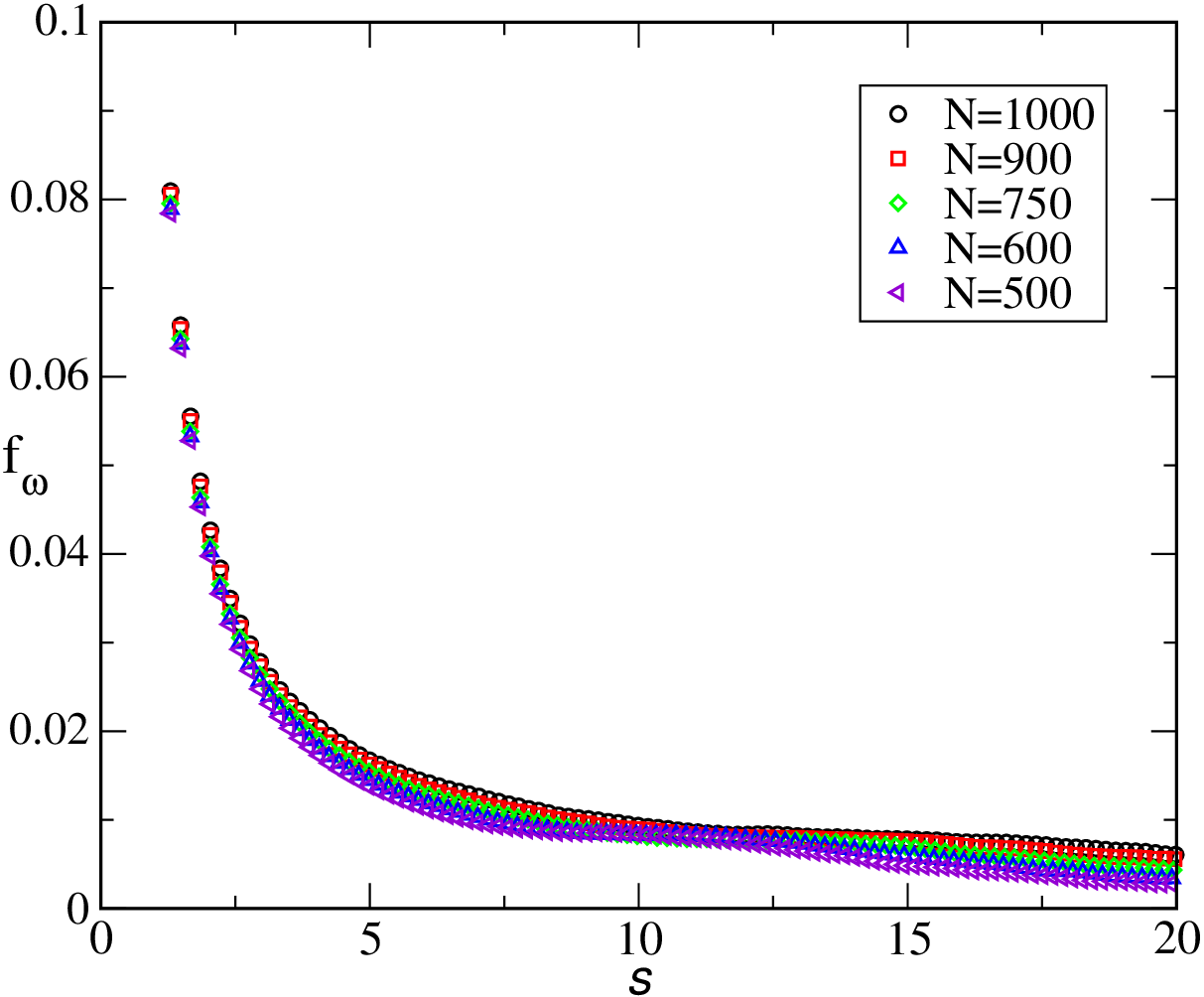}
\caption{(Color online) Dimensionless normalized VACF $f_{\omega}(s)$ for a system of inelastic hard disks in the HCS.  Time $s$ is measured in units of $\left(\widetilde{T}(0)/m \sigma^{2} \right)^{1/2}$, where $\widetilde{T}(0)$ is the initial granular temperature of the system. The coefficient of normal restitution is $\alpha=0.99$ and the density is $n \sigma^{2}=0.385$. Different numbers of particles (and sizes of the system) have been used in the simulations, as indicated in the inset. }
\label{fig1}
\end{figure}

\begin{figure}
\includegraphics[scale=0.6,angle=0]{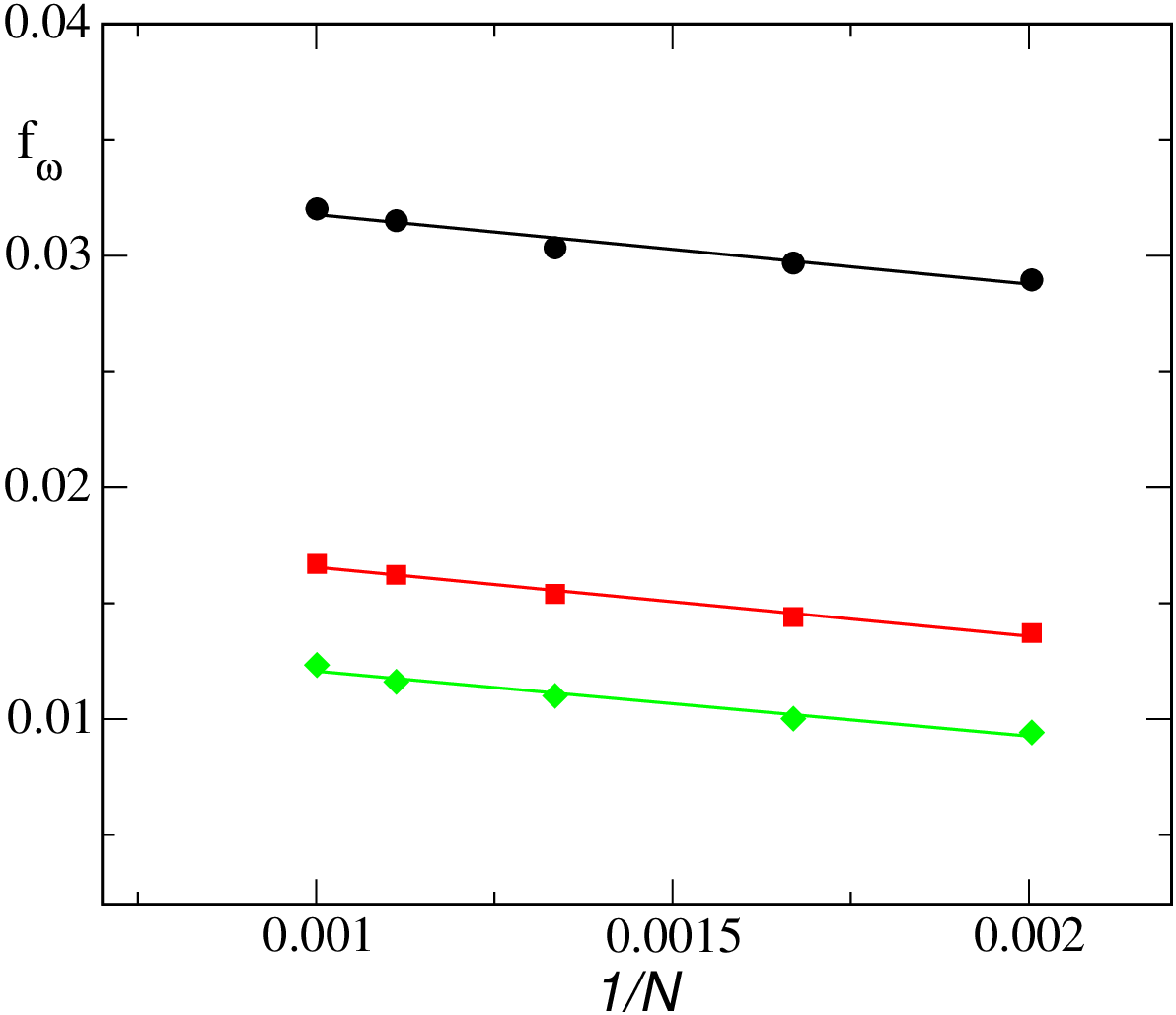}
\caption{(Color online) Values of the dimensionless normalized VACF $f_{\omega}$ as a function of the inverse of the number of particles $N$ of the system. The coefficient of normal restitution is $\alpha=0.99$ and the density $n \sigma^{2} =0.385 $. The symbols are simulation results at three different times, namely $s=2.6$ (black), $s=5.0$ (red), and $s=7.2$ (green), from top to bottom. Time $s$ is measured in units of $\left(\widetilde{T}(0)/m \sigma^{2} \right)^{1/2}$, where $\widetilde{T}(0)$ is the initial granular temperature of the system. The straight lines are  linear fits of the simulation data. The values of the slopes of the three lines are very close. }
\label{fig2}
\end{figure}

In Fig. \ref{fig1}, the results obtained for the decay of the VACF in a system with density $n \sigma^{2}=0.385$ and different values of the number of particles $N$ are shown. The quantity actually plotted is the dimensionless VACF, $f_{\omega}(s)$, defined as
\begin{equation}
\label{6.1}
f_{\omega}(s) \equiv \frac{m C_{\omega \omega}(s)}{\widetilde{T}_{st}}.
\end{equation}
Although small, the dependence of the results on the number of particles is clearly identified in the figure and, at a given time $s$, the VACF is smaller the smaller the system. To investigate the dependence on $N$, three different times, representative of the hydrodynamic relaxation of the VACF, were considered.  At each of these times, the values of the VACF obtained in systems with different sizes but the same density,  were analyzed. The results are given in Fig.\ \ref{fig2}. The solid lines are  linear fits at each time, the slope of the fitting being very similar for the three times considered, namely  $2.93 \pm 0.08$. This slope is about three times the one predicted and observed in  elastic systems at equilibrium, showing that the dependence on the number of particles is larger for inelastic particles in the HCS than for molecular systems at equilibrium

\begin{figure}
\includegraphics[scale=0.6,angle=0]{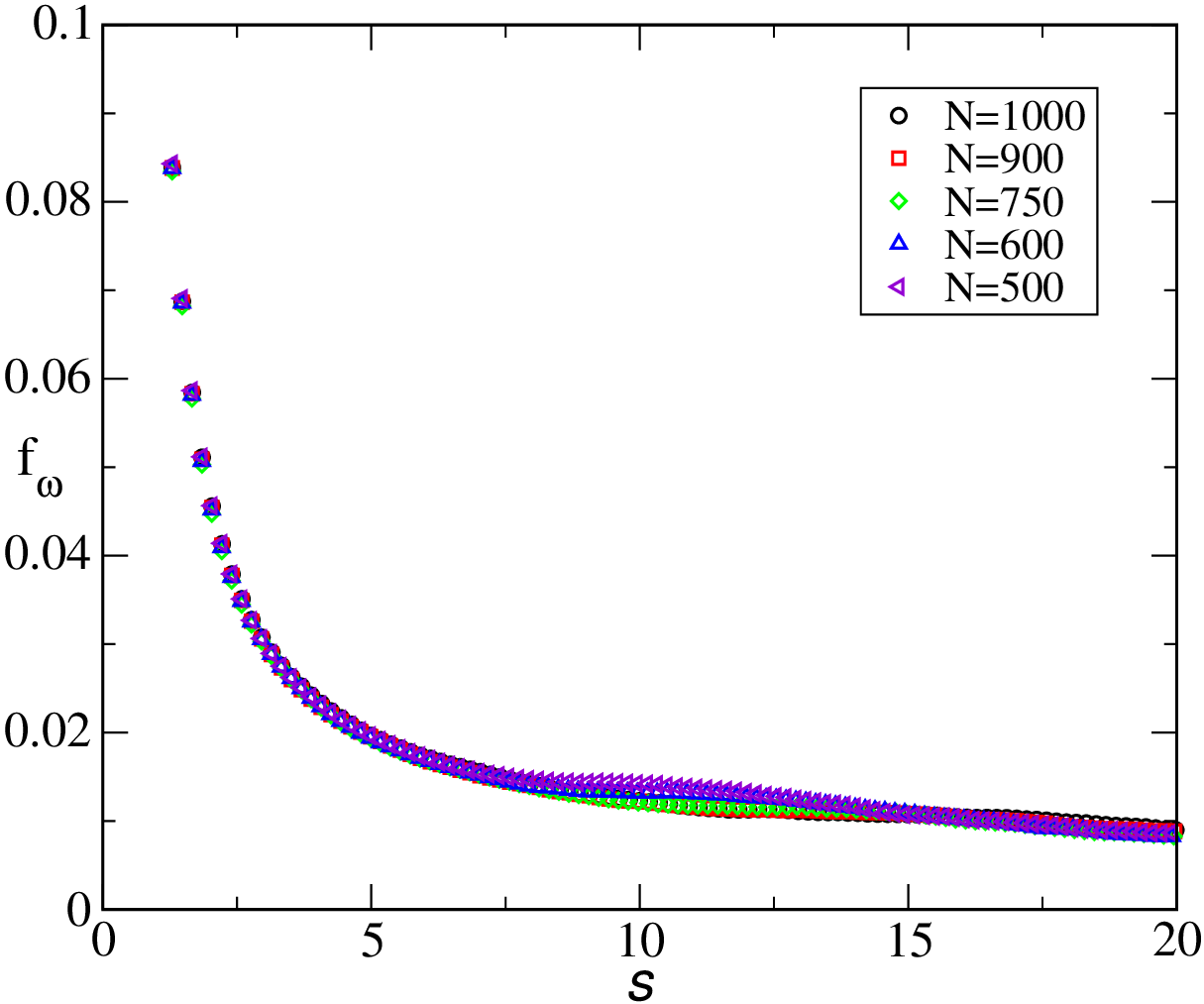}
\caption{(Color online) The same as in Fig.\ \protect{\ref{fig1}}, but now the finite size correction given in Eq.\ (\protect{\ref{6.2}})  has been added to each curve. }
\label{fig3}
\end{figure}

Denoting the VACF measured in the MD simulations with $N$ particles  by $f_{\omega,N} (s)$, it is concluded that the extrapolated value, for an infinite system, is given by
\begin{equation}
\label{6.2}
f_{\omega}(s) \simeq  f_{\omega, N}(s) + 2.93 N^{-1}.
\end{equation}
The modified curves, obtained by adding this correction to the measured VACF reported in Fig. \ref{fig1} are plotted in Fig. \ref{fig3}. It is observed that the collapse of the several curves is very good, especially for $s \lesssim 20 \left( \widetilde{T}(0)/m \sigma^{2} \right)^{1/2}$.  Similar results have been obtained for other densities in the interval considered. Moreover it is found that the prefactor of $N^{-1}$ in the finite size correction term increases as the density increases. On the other hand, in the elastic case, it is always equal to unity, independently of the density.

Before proceeding any further, the above limiting process must be put in a proper context. It is not claimed that the derived results hold in the limit of an infinite system. This is not true, since an infinite system of inelastic hard spheres or disks can not exist in the HCS, because it is very unstable. What has been done is to take into account the conceptual differences between theory and MD simulations, due to the use of different physical conditions as indicated above,  realize that they scale with the inverse of the number of particles, and eliminate those differences to carry out a fair comparison.

\section{The power law tail of the VACF}
\label{s7}
Once it is known how to translate the MD numerical results for the VACF  into results corresponding to the conditions under which the theory is developed, the  existence of a time region in which the VACF presents a power law time tail, as predicted by Eq.\ (\ref{5.8}), will be investigated. As mentioned in the previous section, the observation time in the simulations is limited by the size of the system. Since, once corrected, the resulting VACF does not depend on the number of particles used, only results with the largest number of particles compatible with the stability of the system, namely $N=1000$, will be presented from now on. In any case, only results for  times $s$ shorter than the time it takes a sound wave to cross the system will be shown.

In Fig.\ \ref{fig4}, the scaled VACF of a system of inelastic hard disks  is shown, both as a function of time $s/s_{E}$ and as a function of $(s/s_{E})^{-1}$, for two different densities, $n \sigma^{2}=0.231$ and $n \sigma^{2}=0.385$, respectively. Here
\begin{equation}
\label{7.2}
s_{E}  \equiv \left( \frac{m}{ \pi \widetilde{T}_{st}} \right)^{1/2}\, \frac{1}{2 \sigma n \chi}
\end{equation}
is the mean free time between collisions in the steady representation of the HCS and computed using the Esnskog theory, so that $\chi$ is the equilibrium pair distribution for hard disks at contact. This is a convenient dimensionless time scale to compare results corresponding to different densities, since it is proportional to the cumulative number of collisions per particle which is the relevant time scale for the relaxation of the system. As expected, the velocity correlations are more persistent, i.e. they decay slower, the denser the system. Moreover, a clear region exhibiting a linear dependence on $s^{-1}$ is identified, in qualitative agreement with the theoretical prediction. To carry out a quantitative comparison between the simulation results and Eq.\   ({\ref{5.8}), the latter (with $d=2$) is substituted into Eq.\ (\ref{6.1}), and the result is rewritten in the form
\begin{equation}
\label{7.1}
f_{\omega, \text{hyd}} \simeq \alpha_{D} \left( \frac{s}{s_{E}} \right)^{- 1},
\end{equation}
where the amplitude  $\alpha_{D}$ of the tail is given by
\begin{equation}
\label{7.3} \alpha_{D}= \frac{1}{8 \pi n s_{E} \left( \widetilde{\eta}+ \widetilde{D} \right)}\, .
\end{equation}
 Equation (\ref{7.1}) is of the form considered in the original paper by Alder and Wainwright \cite{AyW70}, and also by other authors \cite{Do75}.

\begin{figure}
\begin{tabular}{c}
\includegraphics[scale=0.5,angle=0]{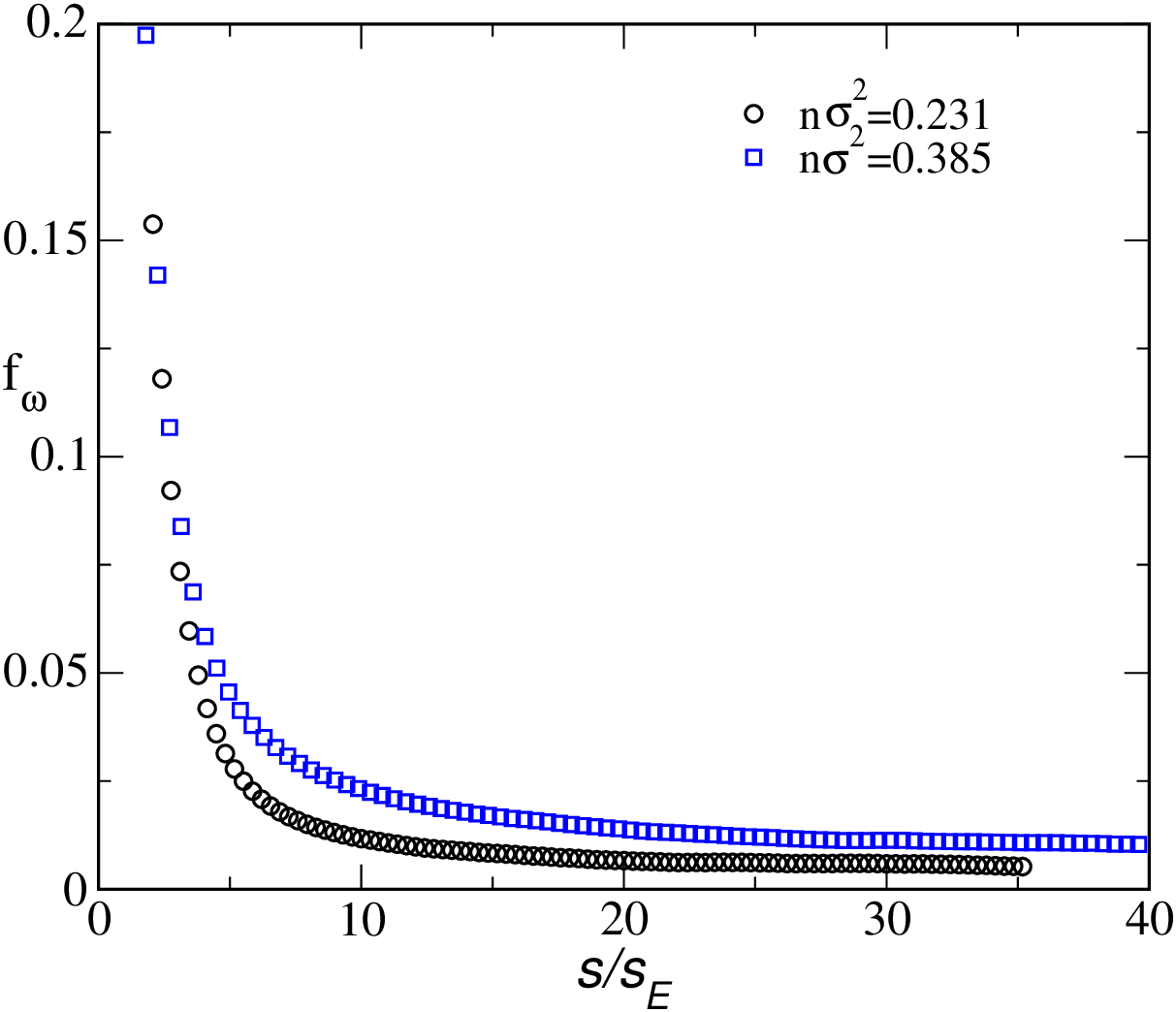} \\
\includegraphics[scale=0.5,angle=0]{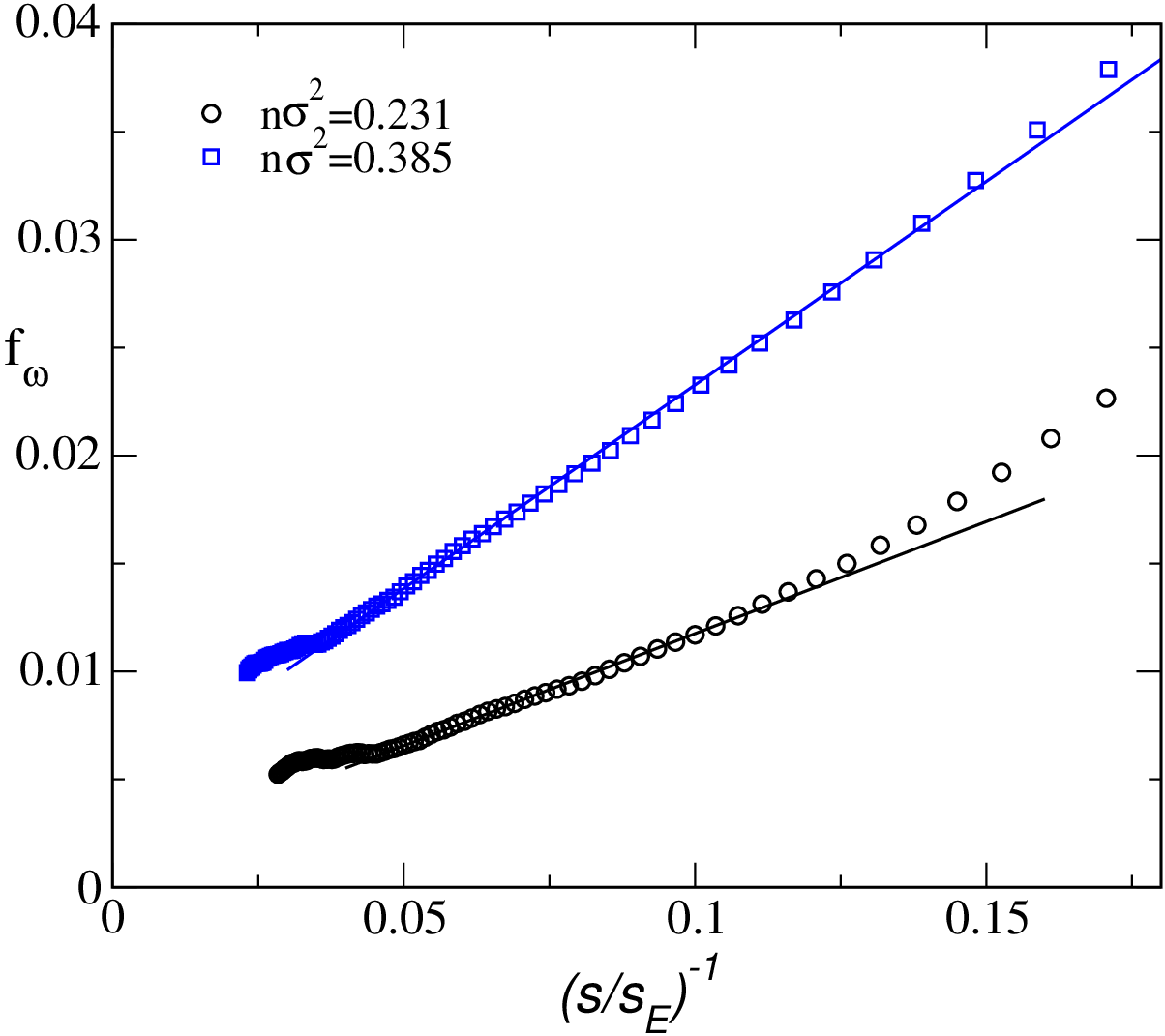}
\end{tabular}
\caption{(Color online) VACF for an inelastic system of hard disks with a coefficient of normal restitution $\alpha=0.99$ and two different values of the density, as indicated in the inset. The results have been obtained in a system with 1000 particles, and finite size effects have been corrected as discussed in the main text. }
\label{fig4}
\end{figure}

The slope of the fits to straight lines of the linear in $(s/s_{E})^{-1}$ regions observed in the decay of the dimensionless VACF $f_{\omega}(s)$ obtained in the MD simulations, provides the values of the tail amplitude. In Fig. \ref{fig5}, $\alpha_{D}$ is plotted as a function of the number density $n \sigma^{2}$ of the system. The symbols are the results from the MD simulations. In addition to the results for inelastic disks with a coefficient of normal restitution $\alpha=0.99$ (red squares), MD values for a system of elastic hard disks (black circles)  have also been included. The latter agree with those reported by Alder and Wainwright many years ago \cite{AyW70,EyW82}. The solid line is the theoretical prediction given by Eq. (\ref{7.3}), using the expressions for the inelastic transport coefficients derived in the Enskog approximation \cite{Lu05,GSyM07}, and the Henderson value \cite{He77} for the pair distribution function at contact. The consistency of using the bare transport coefficients to compute the algebraical tails will be discussed  in the last Section of the paper. For the sake of  completeness, these expressions are reproduced in the Appendix. Actually, the theoretical prediction for the amplitude of the tail for an inelastic system  with $\alpha=0.99$ is indistinguishable from the  theoretical prediction for a system of elastic hard disks. However, the different physical nature of the states considered in both cases must be kept in mind. In particular, it is worth to insist on that a $s^{-1}$ decay of the scaled VACF corresponds to a decay $(t \ln t)^{-1}$ of the VACF in the original time (and velocities) scale.  As already known \cite{Do75}, the agreement between theory and simulations is quite good in the elastic case. For  inelastic systems, the comparison can still be considered as satisfactory, although it is clear that the effects of the inelasticity are much larger than predicted by the theory, The inelastic tail amplitude is up to $15\%$ larger than the elastic value at the highest density considered.

\begin{figure}
\includegraphics[scale=0.6,angle=0]{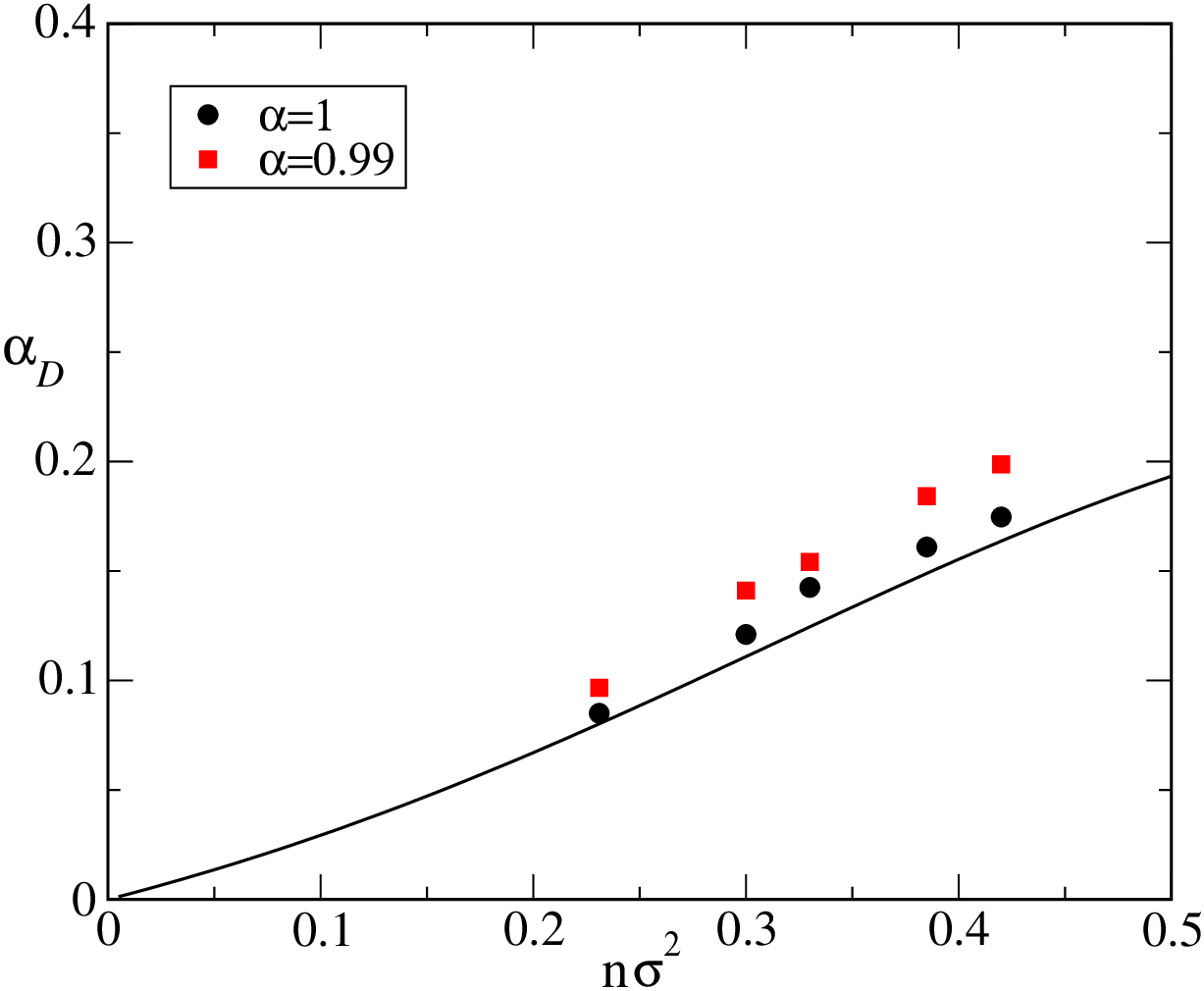}
\caption{(Color online) Tail amplitude $\alpha_{D}$ as a function of the number density of the system. The symbols are simulation results: the (black) circles are for a system of elastic hard disks and the (red) squares for a system of inelastic hard disks with a coefficient of normal restitution $\alpha=0.99$. The solid line is the theoretical prediction using the mode coupling theory developed in the main text and the transport coefficients obtained from the revised Enskog theory. On the scale of the figure, the predictions for the elastic and the inelastic systems are undistinguishable. }
\label{fig5}
\end{figure}

\section{Concluding Remarks}
\label{s8}
Using mode-coupling theory, an approximate expression for the hydrodynamic, long time, part of the relaxation of the VACF of a finite granular gas in the HCS has been derived. The analysis parallels in several aspects the one carried out for elastic systems, but there are many relevant quantitative and conceptual differences. First, the correlation function considered here corresponds to the modified dynamics defined by Eqs.\ (\ref{2.10}) and (\ref{2.11}), in which the particles are submitted to an acceleration, as a consequence of a change in the original time scale. The new time scale measures the average number of collisions per particle, and plays a crucial role for the study of response functions in non-equilibrium systems \cite{DByL02,DyB02,ByR04b}.  The self-diffusion transport coefficient of the granular gas can be expressed in terms of a one-time VACF computed in a steady state only in the modified dynamics. Then, it is the time behavior of the correlation function  in the modified dynamics the one relevant for computing the transport coefficient by means of the Green-Kubo expression.

Another important difference between molecular systems at equilibrium and granular gases in the HCS lies in the meaning of taking the thermodynamic limit in the MD simulations. At a theoretical level, the HCS  can be considered and analyzed at arbitrary size, density, and value of the coefficient of normal restitution, and this legitimates its use as a reference homogeneous state.  On the other hand,  in the simulations a very serious limitation shows up due to the shearing instability.  At fixed density, as the coefficient of normal restitution decreases, the size of the system for which the HCS remains stable decreases rather fast. At moderate  densities, only granular gases of inelastic hard spheres or disks with a very small number of particles can be simulated in the HCS. This limitation is specially severe when the interest is on the long time behavior of a given property, as it is the case here. Actually, it is quite difficult to reach  the time window for which the velocity-autocorrelation function has a power law decay for values of the restitution coefficient not very close to one.

To illustrate the effect of approaching the instability, in Fig.\  \ref{fig6} the VACF of a system in the HCS with $\alpha=0.98$  is shown at three different  times as a function of the inverse of the number of particles $N$. The density of the system is in all cases $n \sigma^{2} = 0.3 $.  For small enough systems ($ N \lesssim 700$),  a linear in $N^{-1}$ fit is accurate, similarly to what happens for $\alpha=0,99$ and $n \sigma^{2}=0.385 $  in Fig. \ref{fig2}.  The measured slope now is $5.17 \pm 0.69$, which is two times the value for $\alpha=0.99$ and the same density. On the other hand, for $N \gtrsim 700$, a departure from the linear dependence is clearly observed. We expect this behavior to be due to the proximity of the instability and the presence of large fluctuations of the transversal velocity field, as predicted by the factor of $(k^{2}-k_{c}^{2})^{-1}$ in the  addends in Eq. (\ref{4.18}).  All the results reported in the figure correspond to states in which all the hydrodynamic fields were observed to stay homogeneous. Therefore, it seems that the hydrodynamic part of the VACF has a divergent behavior as the shearing instability is approached. The same must happen with the (apparent) self-diffusion coefficient.

\begin{figure}
\includegraphics[scale=0.6,angle=0]{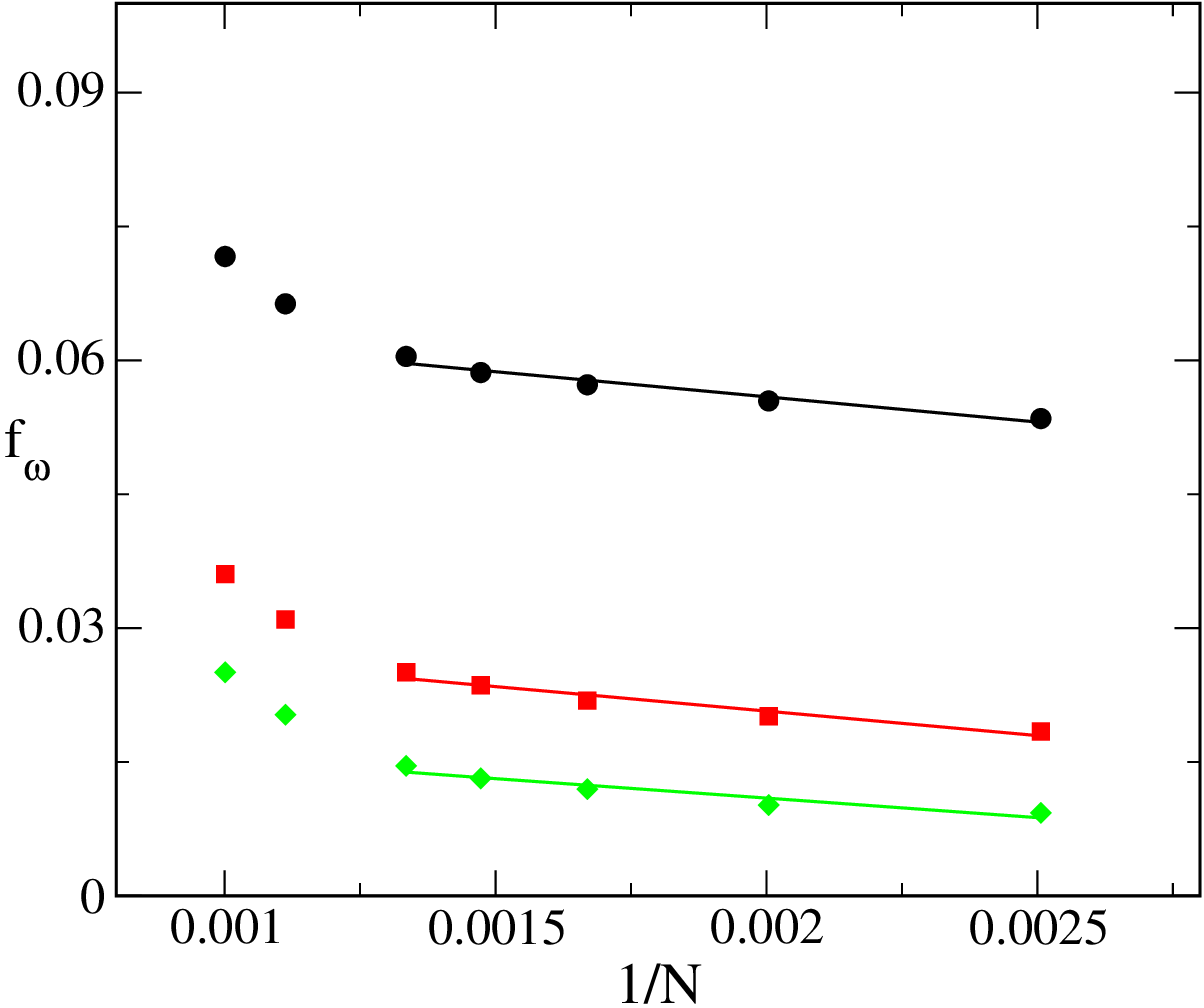}
\caption{(Color online) Values of the dimensionless normalized VACF $f_{\omega}$ as a function of the inverse of the number of particles $N$ of the system. The coefficient of normal restitution is $\alpha=0.98$ and the density $n \sigma^{2}= 0.3 $. The symbols are simulation results at three different times, namely $s=2.4 $ (black), $s=5 $ (red), and $10$ (green), from top to bottom. Time $s$ is measured in units of $\left(\widetilde{T}(0)/m \sigma^{2} \right)^{1/2}$, where $\widetilde{T}(0)$ is the initial granular temperature of the system. The straight lines are  linear fits of the simulation data in the region where this behavior is observed. The values of the slopes of the three lines are very close. }
\label{fig6}
\end{figure}

In two-dimensional molecular systems, it was realized that the analysis similar to the one reported in this paper, is internally inconsistent. The reason is that it is assumed in the derivation that a finite self-diffusion constant $D$ exists. But the asymptotic $t^{-1}$ long-time tail of the VACF implies that the Green-Kubo expression for the self-diffusion coefficient diverges as $\ln t$. It is worth to mention that more refined mode-coupling theories using a time-dependent expression for the self-diffusion coefficient have led to the prediction that the $t^{-1}$ decay corresponds to intermediate times, while a slightly faster decay, namely as $\left( t \sqrt{\ln t} \right)^{-1}$ is expected at later times. Nevertheless, this has never been confirmed by numerical simulations \cite{AWyG71,FNyS77}. The view adopted here, and consistent with the results reported,  is that the influence of the time tail on the {\em observed} value of the self-diffusion coefficient remains negligible over a time scale going well into the time scale in which the $s^{-1}$ tails can be observed. On this scale, $D$, given by the Green-Kubo expression,  appears as constant, the contribution from the tails remaining very small. What happens at the far end of the region in which the self-diffusion coefficient seems to be constant, remains an open question, except by the mentioned theoretical predictions.

A completely different question is the influence of the clustering instability on the {\em observed} self-diffusion coefficient in a finite system  as the shearing instability is approached. This is a very interesting issue that deserves further investigation.

\section{Acknowledgments}

This research was supported by the Ministerio de Econom\'{\i}a y Competitividad  (Spain) through Grant No. FIS2014-53808-P (partially financed by FEDER funds).

\appendix

\section{The cooling rate and the transport coefficients in the Enskog theory }

In this Appendix, the expressions for the cooling rate and the transport coefficients of a system of inelastic
hard spheres or disks of mass $m$, diameter $\sigma$ and constant coefficient of restitution $\alpha$,
obtained by using the Enskog approximation \cite{GSyM07}
 are given for the sake of completeness. The cooling rate $\zeta$ is
\begin{equation}
\zeta=\zeta^{*} \frac{n T}{\eta_{0}}\, ,
\end{equation}
where $\eta_{0}$ is the elastic value of the shear viscosity in the dilute limit,
\begin{equation}
\eta_{0}=\frac{(d+2)\,\Gamma(d/2)}{8 \pi^{(d-1)/2}}(mT)^{1/2} \sigma^{-(d-1)} \, .
\end{equation}
The reduced cooling rate   $\zeta^{*}$ is given by
\begin{equation}
\zeta^{*}=\chi \frac{d+2}{4 d} (1-\alpha^{2}) \left(1+\frac{3}{16} a_{2}\right)\, ,
\end{equation}
with $a_{2}$  the first coefficient of the Sonine expansion of the HCS distribution \cite{GyS95,vNyE98},
\begin{equation}
a_{2}=\frac{16 (1-\alpha)(1-2\alpha^{2})}{9+24 d+(8d-41)\alpha+30 (1-\alpha)\alpha^{2}}\, .
\end{equation}
The shear viscosity $\eta$ is
\begin{equation}
\eta=\eta^{k} \left[1+\frac{2^{d-1}}{d+2} \phi \chi (1+\alpha)\right] +\frac{d}{d+2} \gamma\, ,
\end{equation}
where $\eta^{k}$ is the kinetic contribution to the viscosity,
\begin{equation}
\eta^{k}=\eta_{0} \frac{1}{\nu^{*}_{\eta}-\frac{1}{2}\zeta^{*}} \left[ 1-\frac{2^{d-2}}{d+2} (1+\alpha)(1-3\alpha)\phi\chi\right] \, ,
\end{equation}
while $\gamma$ is the bulk viscosity, that vanishes in the dilute limit,
\begin{equation}
\gamma=\eta_{0} \frac{2^{2d+1}}{(d+2)\pi} \phi^{2}\chi (1+\alpha) \left(1-\frac{1}{16} a_{2}\right)  \, .
\end{equation}
In the above expressions, $\nu^{*}_{\eta}$ is
\begin{equation}
\nu^{*}_{\eta}=\chi \frac{3}{4 d} \left(1-\alpha+\frac{2 d}{3}\right) (1+\alpha) \left(1-\frac{a_{2}}{32}\right)\, ,
\end{equation}
and $\phi$ is the volume fraction,
\begin{equation}
\phi=\frac{\pi^{d/2}}{2^d \Gamma\left(1+\frac{d}{2}\right)} n\sigma^{d}\, .
\end{equation}
Finally, the bare self-diffusion coefficient is given by \cite{BRCyG00}}
\begin{equation}
D=\frac{4}{(1+\alpha)^{2} \left(1+\frac{3 a_{2}}{16}\right)} \frac{\Gamma(d/2) d}{4 \pi^{(d-1)/2} n \chi \sigma^{d-1}} \left(\frac{T}{m}\right)^{1/2}\, .
\end{equation}

In the particular case of hard disks, $d=2$, the pair distribution function at contact $\chi$ is \cite{He77}
\begin{equation}
\chi=\frac{1}{(1-\phi)^2}\left(1-\frac{7\phi}{16}\right)\, .
\end{equation}

\end{document}